\documentclass[conference]{IEEEtran}
\usepackage{graphicx,cite,epsfig}
\usepackage{amsmath} 
\usepackage{amssymb}  
\DeclareMathOperator *{\argmin}{argmin}
\usepackage{pifont}
\usepackage{stmaryrd}
\usepackage{bbding}
\usepackage{subfigure}
\usepackage{algorithm}
\usepackage{algorithmic}
\usepackage{array}
\usepackage{amsthm}
\usepackage{multirow}
\usepackage{url}
\usepackage{mathrsfs}
\usepackage{color, soul} 
\usepackage{mathrsfs}
\usepackage{float}

\usepackage{booktabs}
\usepackage{threeparttable}
\usepackage{amssymb}

\begin{document}
\title{A unified polar decoder platform for low-power and low-cost devices}

\author{\IEEEauthorblockN{Jiajie Tong, Qifan Zhang, Huazi Zhang, Rong Li, Jun Wang, Wen Tong}

\IEEEauthorblockA{Huawei Technologies Co. Ltd.\\
Email: \{tongjiajie, Qifan.Zhang, zhanghuazi, lirongone.li, justin.wangjun, tongwen\}@huawei.com}
}

\maketitle
\thispagestyle{empty}

\begin{abstract}
In this paper, we design a polar decoding platform for diverse application scenarios that require low-cost and low-power communications. Specifically, prevalent polar decoders such as successive cancellation (SC), SC-list (SCL) and Fano decoders are all supported under the same architecture. Unlike high-throughput or low-latency decoders that promote parallelism, this architecture promotes serialization by repeatedly calling a ``sub-process'' that is executed by a core module. The resulting serial SCL-8 decoder is only 3 times as big as an SC decoder. Cost and power are minimized through resource sharing and adaptive decoding techniques, etc. We carried out performance simulation and hardware implementation to evaluate the actual chip area and energy consumption.
\end{abstract}

\section{Introduction}\label{section_introductions}
\subsection{Motivations}
$6$G will cover a wide range of terminals for artificial intelligence (AI) and sensing applications. Among them, the number of the low-end devices is expected to take up the majority. Meanwhile, these numerous devices come in different types, since they need to be customized for thousands of use cases. This poses a challenge for channel coding, as a variety of decoders with different decoding algorithms and parameters are to be implemented.

First, ``versatility'' is required to meet the diverse requirements of application scenarios. However, it is not a wise choice to design a separate type of codes for each scenario due to the overall description and implementation cost. One unified coding scheme would bring a lot of convenience. This in turn requires the decoders in each scenario to be highly specialized and flexibly customized at the same time. To this end, a unified and flexible decoding architecture is desirable.

Second, ``low cost'' and ``low power consumption'' are necessary features for a great proportion of the devices that are small-sized, battery-powered (sometimes even passive devices) with limited hardware fabrication budget. Some devices may transmit several kilobytes per day and receive even less. They can be idle for $99\%$ of the time. Some applications are not delay-sensitive, and the decoding latency requirement can be relaxed. Therefore, the constraint is mainly on the chip size and total energy consumption rather than throughput and energy per bit.

\subsection{Background}
It has been shown in literature that polar codes are versatile and support energy-efficient decoding algorithms. Successive cancellation (SC) is the among the simplest in many soft decoders. It is suitable for resource-constrained hardware but delivers mediocre error correction performance. Successive cancellation list (SCL) decoding runs a list of SC decoder instances in parallel, and requires list management and cross bar, thus is more complex than an SC decoder, although the performance is significantly better. We need to find an architecture that enjoys the benefits of both SC (efficient hardware implementation) and SCL (performance gain).

Polar codes are also versatile on the encoding side. There exist a rich selection of code constructions such as CRC-aided (CA)-polar\cite{CAPOLAR}, parity-check (PC)-polar\cite{PCPOLAR}, and polarization-adjusted polar (PAC) codes\cite{PAC}. They can be unified under the framework of pre-transformed polar codes. They share the same polar transformation module in the encoding part, and most of the decoding modules. Their only differences are their outer codes. Together, they can cover a wide range of use cases, whereas PAC can be optimized for very short block length, and PC-Polar and CA-Polar are designed for longer codes.

In this paper, we focus more on the polar decoding schemes for $6$G low-cost and low-power applications.

\subsection{Contribution}
We designed a unified polar decoder platform to achieve versatility, low power and low cost at the same time. The same SC-decoder-based architecture can be tailored into different chip sizes, and its parameters flexibly configured for different uses. First, its architecture is optimized for our purposes. Second, most of its core modules can be reused across all applications.

Unlike high-throughput or low-latency applications, we can resort to ``serialization'' instead of ``parallelization'' when it contributes to a smaller chip size or lower energy consumption. However, certain parallelism can be preserved as long as they are ``free'', i.e., do not incur additional cost. With serialized implementation, some resource-consuming modules such as cross bar are no longer required.

The technical contributions are summarized:
\begin{enumerate}
  \item The recursive nature of SC decoding algorithm is captured by the ``sub-processes'' in the hardware design. It constitutes the core part of the unified architecture.
  \item With the unified architecture, we implemented three decoders, i.e., SC decoder, adaptive serial-SCL (S-SCL), and Fano decoder. The serial-SCL serializes the list of SC decoder instances and thus simplifies the list management part; In Fano, we apply the multi-bit decision technique and limit the number of tracing attempts.
  \item We evaluated the chip area and energy efficiency based on hardware implementation. With TSMC 16nm process, the sizes of these decoders are $70\mu m\times80\mu m$ for SC, $120\mu m\times140\mu m$ for adaptive S-SCL and $125\mu m\times145\mu m$ for Fano. The power consumption for decoding a packet per second are as follows: $0.326nW$ for SC, $0.553nW$ for adaptive S-SCL, and $41.86nW$ for Fano.
\end{enumerate}

\section{Preliminaries}
\begin{figure}
\centering
\includegraphics[width=0.4\textwidth]{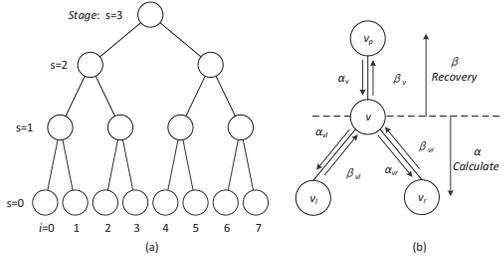} %
\caption{(a) Decoding architecture as a binary tree; (b) Node $\textit{v}$ received/response information}
\label{tree}
\end{figure}

Successive cancellation decoding can be represented as a binary tree traversal\cite{SSC}, as shown in Fig.~\ref{tree}(a). Each subtree therein represents a shorter polar code.
The set of nodes of the subtree rooted at node $\textit{v}$ is denoted by $V_{v}$. Thus $V_{root}$ denotes the full binary decoding tree.
The set of all leaf nodes is denoted by $U$.
Meanwhile, the set of the leaf nodes in subtree $V_{v}$ is denoted by $U_v$. All leaf nodes can be separated into two subsets, one is for information leaf nodes and the other is for frozen leaf nodes.

As shown in Fig.~\ref{tree}(b), a node $v$ in a tree is directly connected to a parent node $p_{v}$, a left child node $v_{l}$ and a right child node $v_{r}$, respectively.
The stage $s$ of a node $v$ is defined by the number of edges between node $v$ and its nearest leaf node. All leaf nodes are at stage $s=0$.

The node $v$, which is not a leaf node, receives $\alpha_{v}$ from its parent node, and generates $\alpha_{v_l}$ according to \eqref{equ:f} \cite{Dec:LLR_based_SCL}.
\begin{equation}\label{equ:f}
f_{-}:\alpha_{v_l}^i = \alpha_{v}^{i}\boxplus \alpha_{v}^{i+2^{s-1}}, i\in[0,2^{s-1}-1].
\end{equation}

Node $v$ sends the $\alpha_{v_l}$ to its left child node and then waits for the feedback vector $\beta_{v_l}$ to return. Subsequently, \eqref{equ:g} \cite{Dec:LLR_based_SCL} is used to calculate $\alpha_{v_r}$ from $\alpha_{v}$ and feedback vector $\beta_{v_l}$.
\begin{equation}\label{equ:g}
f_{+}:\alpha_{v_r}^i = (-1)^{\beta_{v_l}^{i}}\times\alpha_{v}^{i}+\alpha_{v}^{i+2^{s-1}}, i\in[0,2^{s-1}-1].
\end{equation}

After receiving the feedback vector $\beta_{v_r}$ from the right child, node $v$ uses \eqref{equ:psum} to recover the feedback vector $\beta_{v}$, which is sent to its parent node $v_p$.
\begin{equation}\label{equ:psum}
	\left\{
	\begin{array}{lcl}
    \beta_{v}^{i} = \beta_{v_l}^{i}\oplus \beta_{v_r}^{i} \\
	\beta_{v}^{i+2^{s-1}} = \beta_{v_r}^{i}\\
	\end{array} \right.
    , i \in [0,2^{s-1}-1]
\end{equation}

The node $v$, which is a leaf node, receives $\alpha_{v}$ from its parent node, and makes hard bit decision to get the feedback $\beta_{v}$ directly. Thus, a leaf node is a bit-decision node.

Both SC and SCL decoders can benefit from a series of multi-bit decision techniques to prune certain decoding tree branches. Readers may refer to \cite{SSC}\cite{ML}\cite{FSSC}\cite{4T} for several well-known multi-bit decision techniques for SC decoding, and \cite{FSSCL}\cite{GOOD}\cite{SYNDROME} for SCL decoding.
Thanks to these techniques, a none-leaf node can become a bit-decision node because its child leaf nodes no longer need to be visited.

\section{An architecture based on ``sub-process''}
This paper presents a new architecture for polar decoding based on ``sub-process''. SC-based decoding is a typical recursive algorithm. For software implementation, a decoding program runs by recursively calling a subroutine, i.e., decoding a node in Fig.~\ref{tree}(a). Similarly, for hardware implementation, one dedicated module processes a part of the decoding tree at a time, until the decoding tree is fully traversed. The processing is called a ``sub-process'' (SP). The module is called an SP module.

What remains to be designed is how do we partition a decoding tree into SPs, and what exactly constitutes an SP?

Two design principles are crucial for our purpose:
\begin{itemize}
  \item In order to achieve versatility, an SP module should be capable of processing all the SPs in a polar code.
  \item In order to reduce hardware cost and energy consumption, an SP module should be as small as possible.
\end{itemize}

\begin{figure}
\centering
\includegraphics[width=0.4\textwidth]{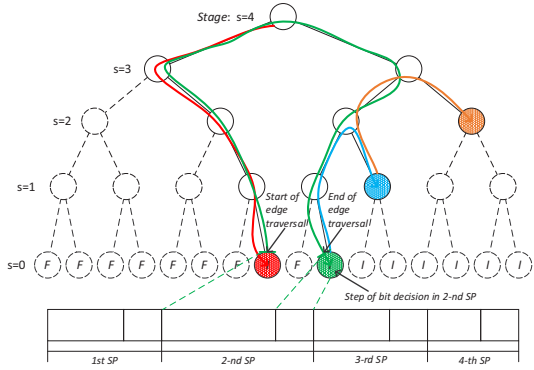} %
\caption{Sub-process in an SC Decoder.}
\label{SP}
\end{figure}

According to the design principles, SP module has to be general enough to process every part of the tree, and at the same time minimizes hardware resource. In other words, it needs to include the components to perform all the decoding functions; but for each function, we can only afford to implement one instance of the component. Correspondingly, we partition a full binary tree such that each part starts from a bit-decision node (or the root node) to the next bit-decision node, excluding the former and including the latter. An example for code length is $N=16$, and code rate $R=0.5$ is given in Fig.~\ref{SP}, where ``\emph{F}'' denotes a frozen leaf node and ``\emph{I}'' denotes an information leaf node. There are four bit-decision nodes in the binary tree (excluding frozen nodes), therefore, the decoding tree can be partitioned into four SPs. As such, each part can be efficiently processed by an SP module with one pass.

Thus, a hardware decoder is composed of an SP module, storage for $\alpha$ and $\beta$, and the necessary control logic. A finite-state machine (FSM) is implemented to repeatedly call the SPs. The SP-based decoding procedures for SC, SCL and Fano are described in the following subsections.

\subsection{``Sub-process'' for SC decoding}
For SC decoding, an SP comprises two steps, ``edge traversal'' and ``bit decision''.
Specifically, an ``edge traversal'' step calculates $\alpha_{v}$ and recovers $\beta_{v}$ for each none-bit-decision node, and a ``bit decision'' step makes hard decisions from the soft input $\alpha$.

Note that the latency of each SP may be different, it depends on the starting node and ending node positions on the tree, and the type of multi-bit decision to be executed at the ending node.

\subsection{``Sub-processes'' for SCL decoding}
The additional functions of SCL over SC decoding are related to list path management. Accordingly, the SP for SCL decoding should additionally include path management related components. Some preliminaries are introduced before defining the SP for SCL, as follows.

PC-SCL and CA-SCL decoders share most of the path management functions except for final path selection.
For PC-SCL decoders, the final path selection is similar to that of SCL. For CA-SCL decoders, there is an additional parameter called ``check times'', i.e., the maximum number of paths to go through CRC check at the end of list decoding.

We use ``SCL$a$T$b$'' to denote a CA-SCL decoder which has list size ``$a$'' and checks the best ``$b$'' paths. For example, SCL8T8 has $8$ list paths, and checks all $8$ paths. Compared with an SC decoder, SCL8T8 has over $1dB$ performance gain in terms of block error rate (BLER) at short code length, e.g., $N=256$. But a typical SCL8T8 requires over ``$8\times$''SC chip area and power consumption.

SCL decoder keeps $L$ survival paths and extends each survival path into two paths once the decoding algorithm visits an information leaf node. A path metric (PM) must be stored to indicate reliability of each extended path. A sorter is used to select the best $L$ survivors from the $2\times L$ extended paths according their PMs. On the other hand, a frozen leaf node only calculates PM.

We adopt the approach of ``good bit'' \cite{GOOD} to simplify SCL decoding. Specifically, a set of very reliable information bits, or good bits, are identified offline.
The SCL decoding algorithm does not perform path extension and sorting for these good bits.

If all leaves in the subtree $V_v$ rooted from a node $v$ are good bits, this node $v$ is called a ``good node'', denoted by $v_G$. Meanwhile, if all leaves in the subtree $V_v$ rooted from a node $v$ are frozen leaves, this frozen node $v$ is denoted by $v_F$.

For a SCL decoding, we identify two types of ``sub-processes'' in one SP module.
The first type is called ``simplified sub-process'' (SSP), whose ending node is either $v_G$ or $v_F$. The steps of SSP are identical to the SP of SC. The second one is called ``full sub-process'' (FSP), which has all components of SSP and additionally includes path management components.

There are four steps in FSP. Besides ``edge traversal'' and ``bit-decision'', FSP requires two more steps, that is, ``sorting'' and ``inter-path data switching''. FSPs and SSPs are repeatedly called to complete decoding. As seen, an SSP is nested in an FSP.

\subsection{``Sub-processes'' for Fano decoding}
Fano decoding was proposed for polar codes in \cite{FANO}\cite{PAC}. Like SCL decoding, it also extends one input path to two output paths at an information leaf node. The main difference is here only one path with the smaller PM will be extended first. In our implementation, the extended path proceeds to the next SP, and the other path is pushed into a stack. If the current path fails, the decoder goes back to retrieve a previous one, following the last-in-first-out order, by popping from the stack.

The decoder uses a threshold to determine whether to extend the current path, or roll back to retrieve a previous path. If the current path's PM exceeds the threshold, the decoder will switch to roll-back mode. During the roll-back mode, the PM of each retrieved path will be examined. If it still exceeds the threshold, the path will be discarded and the decoder goes on to retrieve the next one; otherwise the path will be chosen and extended. In other words, the first path whose PM is less than the threshold will be chosen, and its metadata will be recovered for path extension. If such a path cannot be found, the decoder increases the threshold by a pre-defined increment.

The decoding latency incurred by the above procedures may be prohibitive due to the unlimited back and forth between ``roll back'' and ``path extension''. As a countermeasure, we set up a counter to count the times when a PM exceeds the threshold. In such an event, the decoder has to retrace to a previous path, which is called a ``retracing'' event. In the hardware, we limit the retracing times to bound the worst-case latency.

A CRC-aided Fano decoder performs CRC checks upon reaching the last leaf node. It decides whether to roll back, if CRC check fails, or terminate decoding if passes. Like CA-SCL decoding, it also sets a maximum CRC check times $b$. If the maximum is reached, decoding is terminated regardless of CRC check result.

A Fano decoder also has two types of SPs.
Among them, SSP is exactly the same as in an SCL decoder, but FSP is slightly different. Besides the ``edge traversal'' and ``bit-decision'' steps, here FSP requires a ``retrace decision'' step, instead of the ``sorting'' and ``inter-path data switching'' steps. Because a Fano decoder avoids sorting and path data exchanging, it demands less data storage than SCL decoding.

\section{Low power and low cost design}\label{sp}
\subsection{Serial design for SCL}
\begin{figure}
\centering
\includegraphics[width=0.45\textwidth]{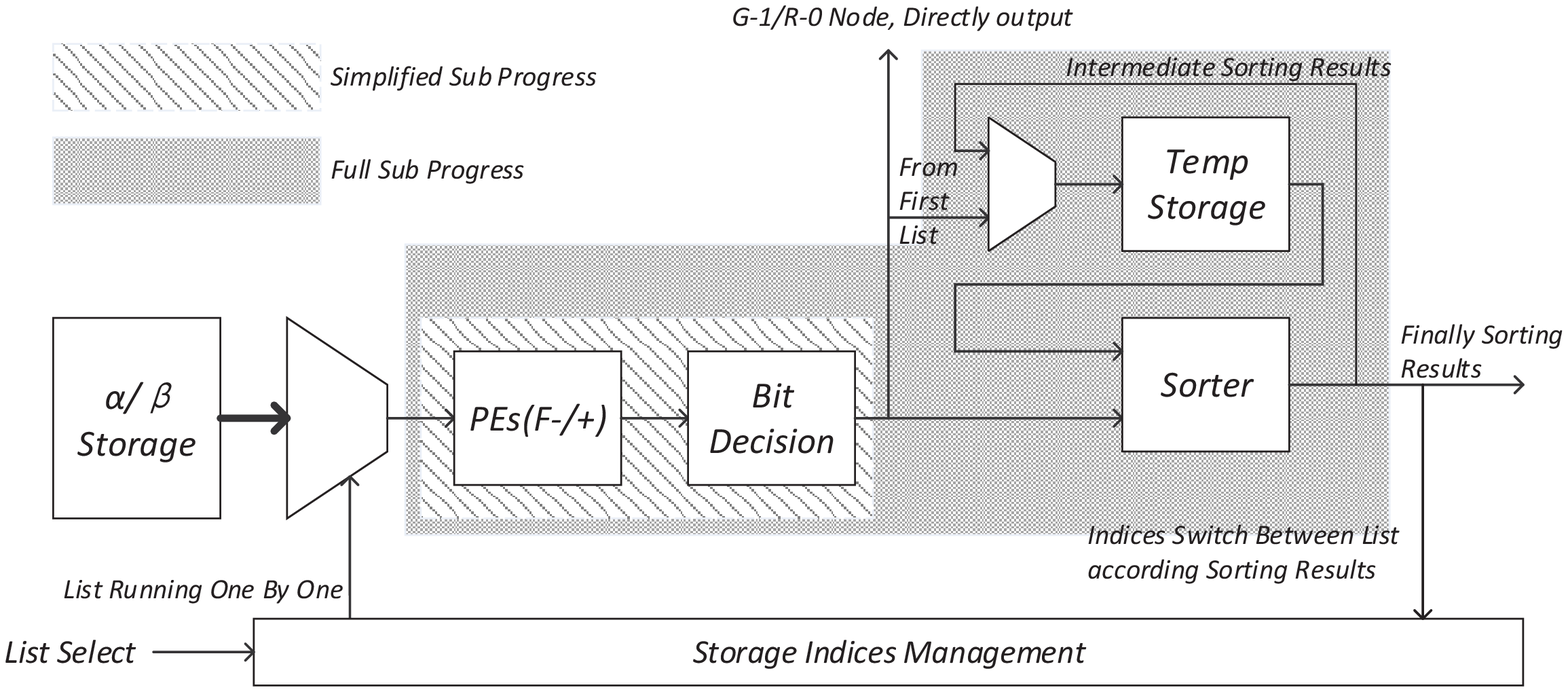} %
\caption{Flow chart of a serial successive cancellation list decoder}
\label{SERIAL}
\end{figure}

To reduce chip area, we design a serial-running architecture for SCL. Fig.~\ref{SERIAL} shows the basic modules of a serial SCL decoder, where list paths are processed one by one. For each path, intermediate LLRs $\alpha$ and partial sums $\beta$ are read from the storage according to addresses provided by an ``index management module''. These variables are subsequently processed by PEs and bit-decision module.

If the current SP is an SSP, it goes on to process the next path. When all paths have been processed, the current SP completes and the decoder proceeds to the next SP. Otherwise if the current SP is an FSP, the PM of a processed path will be sorted together with previous paths' PM (also in a sequential manner). When all paths have been processed, the indices of the surviving (best) paths will go through a small switch network and be sent to the index management module. Then the decoder proceeds to the next SP.

The benefits of a serial architecture are listed below.
\begin{itemize}
  \item All PEs and bit-decision module are shared among paths.
  \item Serialization helps to avoid data switching between the paths, thus no longer requires a big crossbar.
  \item Memories, which is much smaller than registers, are used to save the intermediate results of $\alpha$ and $\beta$, thanks to the low read/write bandwidth requirements.
\end{itemize}
According the evaluation presented in section~\ref{sec:layout}, the layout area of a serial SCL decoder is only $3$ times that of an SC decoder.

\subsection{Adaptive SCL decoding}
An adaptive SCL decoder \cite{adaptive} progressively doubles its list size (e.g., $1 \rightarrow 2 \rightarrow 4 \rightarrow 8$) after a CRC check failure, until a predefined maximum list size is reached. This can effectively reduce every consumption, and its BLER performance matches that of an SCL with the maximum list size.

For hardware implementation, we adopt a simplified version of adaptive SCL. That is, after an SC decoding failure, the decoder directly switches to SCL$8$T$8$ decoding ($1 \rightarrow 8$). We found this strategy saves much controlling overhead and eventually achieves the smallest power consumption over others.

\section{BLER Performance}
\begin{figure}
\centering
\includegraphics[width=0.35\textwidth]{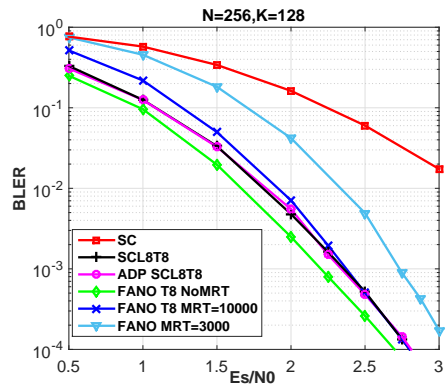} %
\caption{BLER performance comparison.}
\label{BLER}
\end{figure}

\begin{figure}
\centering
\includegraphics[width=0.35\textwidth]{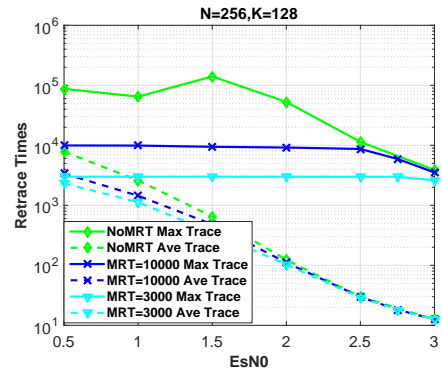} %
\caption{Retracing times for Fano decoding.}
\label{TRACE}
\end{figure}

We compare the BLER performances of various polar decoders for case of code length $N=256$, code rate $R=0.5$. The results are shown in the Fig.\ref{BLER}. Note that MRT in the figure means the maximum retrace times.
The benchmark is the SC decoder, which has the poorest BLER performance.
The adaptive SCL decoder, which is composed of an SC and an SCL8T8, has the same BLER performance of an SCL8T8 decoder. The gain over SC performance is over $1.2dB$ at BLER=$10^{-2}$.

A Fano decoder, which allows a maximum of $8$ CRC check times and unlimited retracing times, has even better BLER performance. However, we observe that retracing times can be as large as $10^5$. Fig.~\ref{TRACE} shows the Fano decoder's worst-case and average retracing times at different SNR values. To bound worst-case latency, we can set the upper limit to 10000 or 3000 and observe their corresponding BLER performance. The former has the a similar BLER performance to SCL8T8 at high SNR but a $0.2dB$ loss at low SNR. The latter has a $0.4dB$ gap from SCL8T8 at all SNRs.

\section{Hardware Implementation}\label{hardware}
\subsection{One platform for all devices}\label{sec:COMODE}
For 6G, a vendor may customize a great amount of decoders for thousands, if not millions, of low-end devices.
Besides the aforementioned techniques to bring down per-device cost, the overall development, fabrication and manufacturing cost can be reduced a ``platform sharing'' strategy. This demands a unified hardware architecture to support all polar decoders.

The SP-based architecture lends itself well to platform sharing. We obverse that the SPs of these decoders have the similar steps. The FSM-based control logic and part of the storage can also be shared across different decoders. These functions are implemented in a core module, and we build around it a unified architecture with peripheral add-on modules. This architecture can be easily reconfigured to different decoders with different parameters.

\begin{figure}
\centering
\includegraphics[width=0.45\textwidth]{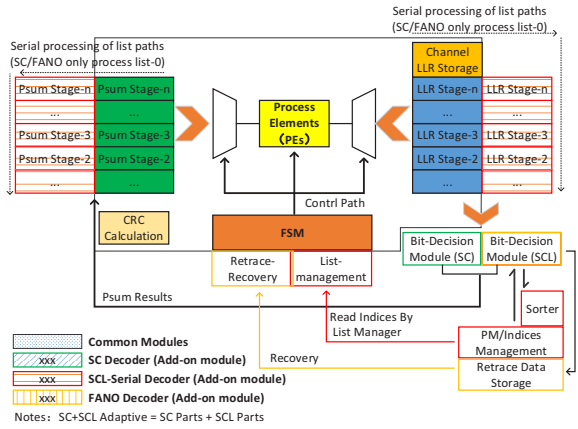} %
\caption{A unified hardware architecture for polar decoders.}
\label{comode-archi}
\end{figure}

The Fig.\ref{comode-archi} shows the unified architecture for polar decoders. The core module includes FSM for SP coordination, storage for channel LLRs, and intermediate LLRs $\alpha$ and partial sums $\beta$, and PEs for $f_{-/+}$ calculation. A list of add-on modules for different types of decoders are described below.
\begin{itemize}
  \item \emph{SC decoder:} bit decision module for SC.
  \item \emph{Serial SCL decoder:}\footnote{SCL decoder was not implemented as an independent decoder.} bit decision module for SCL decoder, additional storage for $\alpha$ and $\beta$, a sorter, and a list path management module (PM/indices).
  \item \emph{Fano decoder:} reuse the bit decision module of SCL; plus additional storage for retracing, and path recovery module for retracing.
  \item \emph{Adaptive decoder:} reuse the add-on modules of SC and serial SCL decoders, plus control logic for triggering SCL decoding.
\end{itemize}

\subsection{Implementation of decoders}\label{sec:IMP}
We designed three types of decoders based on the proposed architecture to verify the area efficiency and energy efficiency.
Two sets of PEs are implemented, the first contains $8$ PEs and the second contains $4$. They are concatenated as described in \cite{Jiajie}. This structure can avoid $\alpha$ storage for stage $s={3,5,7}$.
Multi-bit decisions are made at stage $2\leq s\leq4$.
A CRC check module is implemented for error detection.
All designs employ $6$-bit quantization for LLRs.

\begin{figure}
\centering
\includegraphics[width=0.4\textwidth]{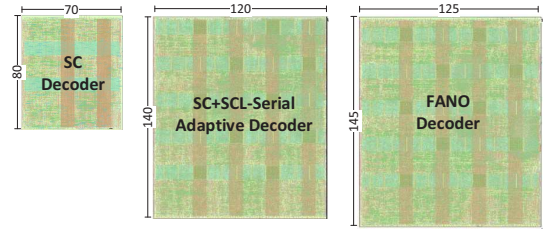} %
\caption{Layout graphs of an SC Decoder, an adaptive (SC+SCL) decoder and a Fano decoder, drawn to the same scale.}
\label{fig:layout}
\end{figure}

\subsubsection{SC decoder}
\begin{figure}
\centering
\includegraphics[width=0.5\textwidth]{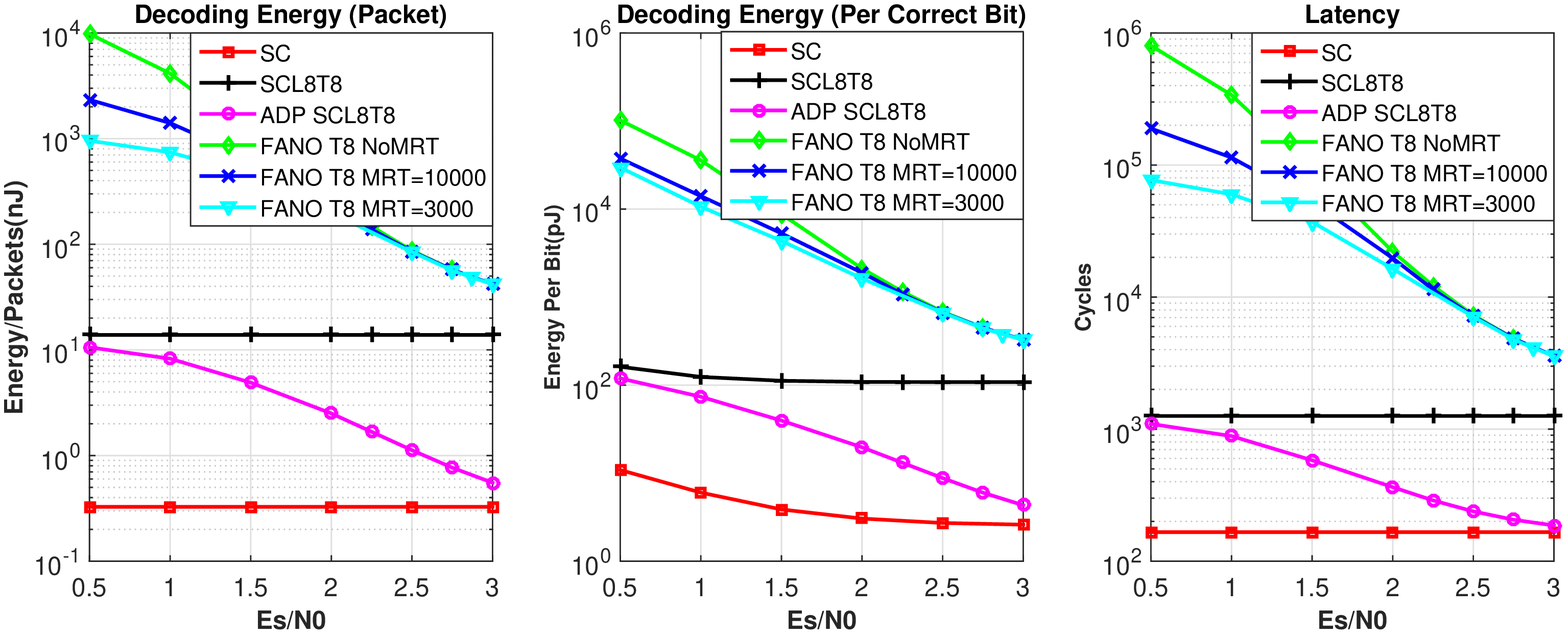} %
\caption{Left: Decoding energy per packet ($nJ/Pkt$). Center: Decoding energy per correct bit (pJ/bit). Right: The decoding latency between each decoders(Cycles).}
\label{power}
\end{figure}

To reduce latency, we employ multi-bit decision for SC. It simplifies the decoding of rate-1 (R-1)\cite{SSC}, single parity check (SPC)\cite{FSSC}, repetition (REP)\cite{FSSC}, dual-SPC (SPC2)\cite{4T}, dual-REP (REP2)\cite{4T}, parity checked repetition (PCR)\cite{4T} and repeated parity check (RPC) \cite{4T} nodes.

\subsubsection{Adaptive SC and SCL$8$T$8$-serial decoder}
Besides the modules in SC, an SCL decoder also includes additional multi-bit decision modules to support flip-syndrome algorithm at stage $s=2$\cite{SYNDROME}, which can avoid the sorting between the candidates extended from the same path. Meanwhile, this module also supports fast decoding at nodes of $v_G$ and $v_F$.
\subsubsection{Fano decoder}
A Fano decoder reuses the same multi-bit decision as in the adaptive decoder. It keeps a maximum of $6$ candidate paths at path extension. The best candidate is selected as the output of the current SP, and the others are pushed back to the first-in-last-out stack.

\subsection{Layout view}\label{sec:layout}
We present the physical implementations of the three decoders with TSMC $16$nm process. The decoding clock frequency for all these decoders is $50Mhz$.
The chip area of these decoders are $70\mu m\times80\mu m$ for SC, $120\mu m\times140\mu m$ for adaptive SCL and $125\mu m\times145\mu m$ for Fano.
Fig.~\ref{fig:layout} demonstrates the layout graphs of them.

Thanks to the serialized and unified architecture, the area of the adaptive SCL with list size $8$ is only $3$ times as big as the SC decoder, much smaller than the conventional implementation that requires at least $8$ times chip area.
The Fano decoder is slightly bigger than the adaptive decoder, mainly resulting from the stack for storing candidate paths. According the synthesis area reports, the area of candidate stack is about $41.3\%$ of the whole Fano decoder.

\section{Key Performance Indicators}\label{KPI}
The key performance indicators (KPIs) are reported in this section are energy consumption per packet, energy consumption per correctly decoded bit, and decoding latency.
We present the indicators in the Fig. \ref{power}.

First of all, we evaluate the average decoding energy per packet.
The evaluation is based on a simulation in which $200$ packets are decoded. The tested polar codes have code length N=$256$ and code rate R=$0.5$.
For an SC decoder, it takes about $0.3\sim0.4$nJ to decode a packet. Meanwhile, an SCL$8$T$8$ decoder takes $13\sim15$nJ. Both energy consumptions do not change with SNR. An adaptive (SC+SCL8T8) decoder consumes about $11$nJ at low SNR. When ESN0 increases to $3dB$, the energy consumption reduces to only $0.55$nJ. The energy consumption of a Fano decoder exhibits the same trend as the adaptive decoder, but overall is $10^2\sim10^4\times$ higher. At low SNR, a Fano decoder takes $10^4$nJ to decode a packet given unlimited retracing times. Its energy consumption will reduce to $2333$nJ and $955$nJ, if the retracing times are limited to $10^4$ and $3\times10^3$, respectively. Note that at high SNR, the energy consumptions under all three retracing settings are almost the same, at $41\sim43$nJ per packet.

The second indicator is decoding energy for per correct bit. We exclude incorrect bits to avoid trivial solutions (such as not decoding at all). Thus, the energy consumption of SC and SCL$8$T$8$ decreases as SNR increases, as more correct bits are decoded. For an SC decoder, a correct bit costs $10.9$pJ on average at EsN$0=0.5dB$, and $2.6$pJ at EsN$0=3dB$. For an SCL$8$T$8$ decoder, a correct bit costs much more energy: $162$pJ and $108$pJ at EsN$0=0.5dB$ and $3dB$, respectively. An adaptive decoder achieves the same performance as SCL$8$T$8$, but requires much less energy: $119$pJ to $4.3$pJ, comparable to that of an SC decoder at high SNR.
The Fano decoder with different retracing settings are also evaluated. At low SNR, a correct bit comsumes $10^5$pJ with unlimited retracing times, and it reduces to $3.7\times10^4$pJ and $3\times10^4$pJ when retracing times are limited to $10^4$ and $3\times10^3$. At high SNR, all energy consumptions reduce to $325\sim328$pJ regardless of retracing setting.

The last indicator is average decoding latency. Although serialization trades latency for low power consumption, too much decoding time keeps the device awake and eventually consumes more energy. A SC takes $166$ cycles to decode a packet and a SCL$8$T$8$ costs $1258$ cycles. The average latency of adaptive decoding is similar to SCL$8$T$8$ at low SNR region and similar to SC at high SNR. The former is $1096$ and the latter is $186$ cycles. Fano decoder has the maximum decoding latency among these decoders. At low SNR, a packet takes $8\times10^5$ cycles per packet without retracing limitation, and it reduces to $1.9\times10^5$pJ and $7.7\times10^4$pJ, with a retracing limitation of $10^4$ and $3\times10^3$. At high SNR, the decoding latencies reduce to $3570\sim3578$ cycles for all cases.

\section{Conclusions}
\label{section_conclusions}
In this paper, we implement three polar decoders with a unified hardware architecture for \emph{low cost} and \emph{low power consumption} scenarios. This architecture achieves versatility with a core module called ``sub-process'' and add-on modules. The core module is shared among different decoders, and add-on modules support various decoding algorithms. Among them, the SC decoder is only $4100\mu m^2$ and has the lowest cost and power consumption.
An adaptive SC/SCL-8 decoder requires only 3 times area of an SC decoder, and enjoys both benefits of low power and good BLER performance. Finally, Fano decoder without retracing limitation achieves the best BLER performance, still low cost, but incurs the highest energy consumption. They together can serve a wide range of potential applications in 6G.

\end{document}